\documentclass[aps,pre,twocolumn,floatfix,superscriptaddress]{revtex4}

\usepackage{graphicx,psfrag,color}
\usepackage{bm}

\newcommand{\average}[1]{\left<{#1}\right>}

\newcommand{\E}{\mathrm{e}}
\newcommand{\D}{\mathrm{d}}

\newcommand{\omt}{\kappa_T}
\newcommand{\bT}{\overline T}

\newcommand{\p}[1]{\left({#1}\right)}

\newcommand{\pg}[1]{\left\{{#1}\right\}}
\newcommand{\derpart}[2]{\frac{\partial #1}{\partial #2}}

\newcommand{\Po}{P_{out}}

\usepackage{comment}
\includecomment{mycomments}

\newcommand{\commentOut}[1]{}

\begin{document}

\title{Out-of-equilibrium Frenkel-Kontorova model }

\author{A. Imparato}

\affiliation{Department of Physics and Astronomy, University of Aarhus\\ Ny 
Munkegade, Building 1520, DK--8000 Aarhus C, Denmark}

\date{}

\begin{abstract}
A 1D model of interacting particles moving over a periodic substrate and in a position dependent  temperature profile is considered. 
When the substrate and the temperature profile are spatially asymmetric a center-of-mass velocity develops, corresponding to a directed transport of the chain.
This autonomous system can thus transform heath currents into motion.
The model parameters can be tuned such that the particles exhibit a crossover from an ordered configuration on the substrate to a disordered one, the maximal motor effect being reached in such a disordered phase. In this case the manybody motor outperforms the  single motor system, showing the  great importance of collective effects in microscopic thermal devices. Such collective effects represent thus a free resource that can be exploited  to enhance the dynamic and thermodynamic performances in microscopic machines.
\end{abstract}
\maketitle

\section{Introduction}
The Frenkel-Kontorova (FK) model is one of the simplest yet most successful models in low-dimensional condensed matter physics, and provides, despite its elementary form,  
a realistic description of the structure and dynamics of a chain of atoms adsorbed on a periodic substrate \cite{FK38,FK38a,Frank49}.
While its most common application is the description of crystal lattices in presence of dislocations, the model has been used to study dislocation (fluxon) dynamics in Josephsons junctions, the dynamics of domain walls in magnetic systems,
 the nonlinear dynamics of DNA models \cite{Braun04} and even the friction dynamics of sliding surfaces \cite{van_den_Ende_2012,Norell16}. Furthermore, the FK model has been used to study heat transport in classical 1D non-linear systems \cite{Hu98,Lepri03}.

On the experimental side, the FK model has been realized by using chains of ions in optical lattices \cite{Bylinskii2015,Gangloff2015,Bylinskii2016,Kiethe2018}, and there has been a recent proposal to realize the model by using Rydberg atoms \cite{Munoz2020}.

The model, in its simplest form, consists of $N$ interacting ``atoms'', each sitting on a 1D periodic potential with period $b$ (the substrate).
Of particular interest for the present study is the fact that at low temperature, the FK model exhibits a  phase transition between a  commensurate (C) and an incommensurate (IC) phase \cite{Chaikin}.
The term commensurate refers to an ordered phase where the atoms are locked-in to the periodic substrate, forming periodically repeated cells. The simplest commensurate states are those where all atoms sit in a minimum of the substrate, the distance between two successive atoms being a fixed integer multiple of $b$. On the contrary, the regular arrangement of the atoms on the substrate is broken in the IC-phase. We will be more specific in the following, and provide some explicative examples.

Collective effects such as synchronization and  phase transitions have been shown to enhance the performance of interacting thermodynamic machines.

System of microscopic interacting Otto engines can approach the Carnot efficiency at finite output power, when operating in proximity  of a second-order phase transition \cite{Campisi2016}.

In models of interacting
molecular motors undergoing a dynamical phase transition, the efficiency at maximum power in many-motor systems is larger than in the single motor case \cite{Golubeva2012a,Golubeva2013,Golubeva2014}.
In systems of interacting  work-to-work transducers the system dynamical phase can be tuned such that the output power or the efficiency at maximum power turn out to be larger than in the single isolated transducer \cite{Imparato15,Herpich18,Herpich18a}.

Finally, in a 2D system of rotors driven out of equilibrium by a temperature gradient, and acting as a thermal autonomous motors \cite{SuneImparato19},  the dynamical response turns out to be  maximal at the phase transition between an ordered and  a quasiliquid phase.

Thus, in general from the above studies the following conclusion can be drawn: systems of microscopic devices at the verge of an equilibrium or out-of-equilibrium phase transition   operate in an optimal working regime.
Furthermore, the results of refs.~\cite{Campisi2016,SuneImparato19} support the following interpretation: while a thermodynamic system at the verge of a second-order equilibrium phase transition is maximally susceptible to  a change in  generalized forces, when an out-of-equilibrium disturbance is applied in proximity of a phase transition the non-equilibrium response of the system is enhanced, in terms of larger mechanical currents or extracted power.

In view of these considerations, and given that the C-IC transition in the FK model is a genuine collective effect, we will put it to good use  in order to design an efficient many-particle transport mechanism.

Specifically, in the following we will consider a FK model diffusing in a position-dependent and periodic temperature profile, and show that this model exhibits a directed transport when the substrate potential and the temperature profile break a specific spatial symmetry. In particular the maximal velocity is achieved in the IC-phase, for which the $N\to\infty $ model always exhibits a center-of-mass velocity which is larger than the one particle model. 

It is worth mentioning that the mechanism driving the particle transport in the present model is completely different from the one considered, e.g., in \cite{Braun04}, where an external constant or time-dependent force is applied on the system, thus having an external (mechanical) agent directly pumping energy into the FK-chain, and therefore inducing motion in, e.g., tilting, rocked or pulsating ratchets (see \cite{Braun04} and references therein). In contrast to those setups,  the present construction is completely autonomous: Once the substrate potential  and the temperature profile have been set up, the system exhibits directed transport without the intervention of any additional mechanical driving, as detailed below.
In that respect the present setup is akin to the two-temperature autonomous motors that convert heat currents into motion, considered in, e.g., \cite{Fogedby17,Fogedby18,Sune19a,Drewsen19}.


The paper is organized as follows: in section \ref{sec:model} we describe the FK model and the stochastic equation governing its motion.
In section \ref{sec:res} we present the properties of the equilibrium and the out-of-equilibrium FK model subject to thermal noise, in particular we study the steady state velocity in the out-of-equilibrium case. The efficiency of the motor is discussed in section \ref{sec:eff}. We conclude in section \ref{sec:conc}.

\section{The model}
\label{sec:model}

Let us consider the Frenkel-Kontorova model with potential energy
\begin{equation}
U(\pg{x_i})=\sum_{i=1}^{N} \frac K 2 (x_{i+1}-x_i-a)^2+ V(x_i),
\label{U:def}
\end{equation} 
with 
\begin{equation}
V(x)= -V_0 \cos (2 \pi x/b),
\label{ext:pot}
\end{equation} 
and open boundary conditions. 
Let $\tilde a$ be the average distance between two particles,
\begin{equation}
\tilde a=\lim_{N\to\infty}\frac{(x_{N}-x_1)}{N}.
\label{ta:def}
\end{equation} 
At low temperature, where thermal fluctuations are negligible, the model exhibits a transition from a commensurate phase ($\tilde a/b$ rational)
to incommensurate phase ($\tilde a/b$ irrational), which is characterized by {\it discommensurations} (defects) disrupting the order in the chain \cite{Chaikin}, and the relevant parameter that controls the phase of the system is  $a/b$.

Here we consider the case where the FK chain diffuses on a substrate with position dependent and periodic temperature profile. To keep the model simple, in the spirit of its equilibrium counterpart, in the following we consider
\begin{equation}
T(x)=\bT-\Delta T/2 \cos( \omt x+\phi_T).
\label{eqT}
\end{equation} 
 
The monomeric version of the this out-of-equilibrium FK model is known as the B\"uttiker-Landauer (BL) model \cite{Buttiker1987,Landauer1988}:  a single particle diffusing in a periodic potential $V(x)$, with a position-dependent and periodic temperature profile $T(x)$. For the specific case where  $V(x)$ and $T(x)$ have the same period $L$, the single particle exhibits a net non-zero velocity if the effective potential $u(x)=\int_0^{x} \D y\,  V'(y)/T(y)$ is not a periodic function of $x$ \cite{Matsuo2000,Fogedby17}, see also appendix \ref{appa}. In this case the quantity $u(x+L)-u(0)$  amounts to the entropy production upon completing a cycle, with the particle moving a distance $L$ in the positive direction. 

The BL model is thus the monomeric ($N=1$) counterpart of the FK chain considered here.
In comparison to it, the FK model with $N$ interacting particles has an additional parameter, namely the particle-particle equilibrium distance $a$ that, as we will see, plays a fundamental role in determining the system different phases, which in turn are  characterized by different propulsion effects in systems with a relatively large $N$.


The FK can be studied exactly in the limit of vanishing temperature, and for small substrate undulations  as represented by $V_0$ \cite{Chaikin,Braun04}. 
Here a non vanishing temperature profile is needed to support the motor effect, and furthermore  the case of arbitrary substrate undulations will be considered. Therefore we will resort on numerical computer simulations.

Specifically we will integrate the Kramer's equation of motion 
\begin{equation}
m \ddot x_i=-\partial_i U-\gamma \dot x_i + \sqrt{2\gamma T(x_i)}\xi_i(t),
\label{Kram:eq}
\end{equation} 
with $\langle\xi(t)\xi(t')\rangle=\delta(t-t')$, $k_B=1$ and open boundary conditions.
The choice of open boundary conditions is motivated by the fact that we are interested in the long time transport dynamics of the model.
We use the full Kramer's equation and not the overdamped Langevin equation as the latter is known to fail to predict the heat transfer in the BL model \cite{Derenyi99,Hondou2000,Benjamin08}. The evaluation of the heat is indeed required  to calculate the FK model's thermodynamic efficiency, as discussed in section \ref{sec:eff}.


\section{Equilibrium and out-of-equilibrium properties}
\label{sec:res}
\begin{figure}[h]
\center
\psfrag{a}[ct][ct][1.]{$a/b$}
\psfrag{al}[ct][ct][1.]{$(a)$}
\psfrag{a/b}[ct][ct][1.]{$a/b$}
\psfrag{dx}[ct][ct][1.]{$\average{\tilde a}/b$}
\psfrag{V1}[cc][cc][.8]{$V_0=1$}
\psfrag{V5}[ct][ct][.8]{$V_0=5$}
\psfrag{V10}[cc][cc][.8]{$V_0=10$}
\psfrag{d=1}[ct][ct][.8]{$\kappa=1$}
\psfrag{d=2}[ct][ct][.8]{$\kappa=2$}
\psfrag{d=3}[ct][ct][.8]{$\kappa=3$}
\psfrag{IC}[cr][cr][.8]{IC}
\includegraphics[width=8cm]{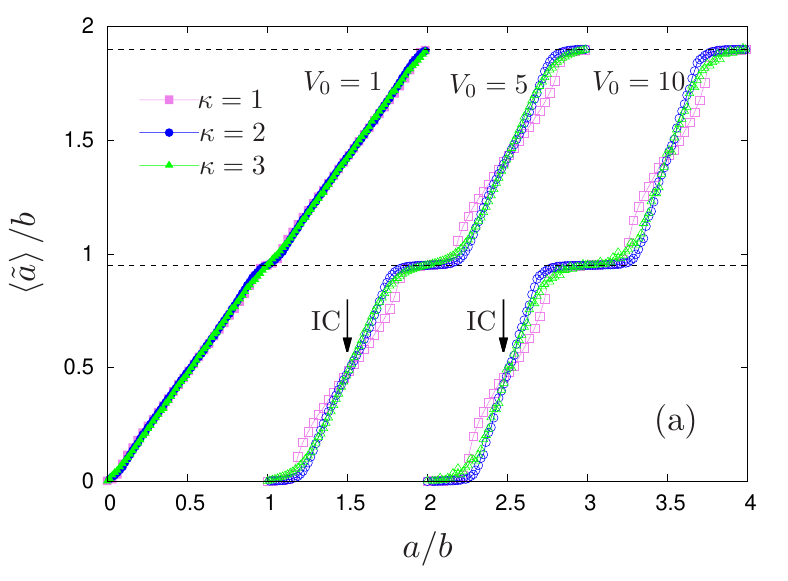}
\psfrag{V10}[ct][ct][.8]{$V_0=10$}
\psfrag{d=2T2}[ct][ct][.7]{$\bT=3.5$}
\psfrag{d=2T1}[ct][ct][.7]{$\bT=1.75$}
\psfrag{T=0.875}[ct][ct][.7]{$\bT=0.875$}
\psfrag{d=2DT0}[ct][ct][.7]{$\Delta T=0$}
\psfrag{DT}[r][r][.7]{$\Delta T=1.5$}
\psfrag{b}[ct][ct][1.]{$(b)$}
\psfrag{c}[ct][ct][1.]{$(c)$}
\includegraphics[width=8cm]{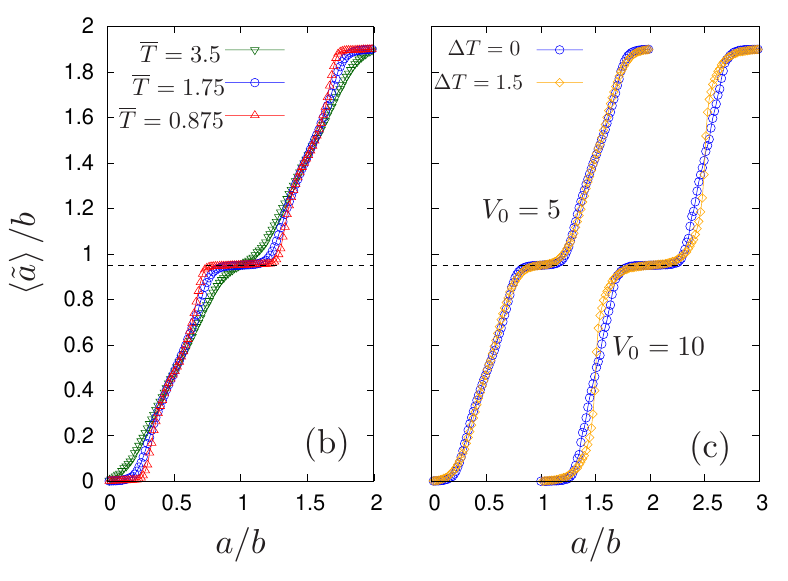}
\caption{(a): rescaled average particle-particle distance $\average{\tilde a}/b$ (\ref{ta:def}) as a function of the rescaled lattice spacing $a/b$ for the equilibrium system $\Delta T=0$ and $\bT=1.75$, $K=5$ and different values of $\kappa=2 \pi/b$ and of the undulation amplitude $V_0$. The periodic curves with $V_0=5$ ($V_0=10$) have been shifted by 1 unit (2 units) along the $x$-axis for clarity. The dashed horizontal  lines correspond to the C-phase where the order parameter $\tilde a/b$ is a multiple integer of $(N - 1)/N$, with $N=20$. The vertical arrows denote the maximally IC-phase.
(b):  $\tilde a/b$ as a function of $a/b$ for different values  of  $\bT$ with $\Delta T=0$,  $K=V_0=5$, $\kappa=2$. (c): comparison between the equilibrium case  $\Delta T=0$ and the out-of-equilibrium case  $\Delta T\neq 0$, for two values of $V_0$, with $\bT=1.75$, $K=5$, $\kappa=\kappa_T=2$, $\phi_T=\pi/2$. The set of curves with $V_0=10$ is shifted along the $x$-axis for clarity. All the curves are obtained through numerical integration of eq.~(\ref{Kram:eq}) with $m=0.01$, $\gamma=1$,  $10^3$ independent trajectories and $10^3$ time steps. We will use these four values in the other figures, unless differently stated. } 
\label{ord:fig}
\end{figure}

Let us first consider the equilibrium properties of the FK model with $\Delta T=0$.
The relevant parameter to study is the end-to-end length $\tilde a$ as a function of the inter-particle equilibrium distance $a$.
The two limiting cases, namely $V_0\to0$ and $V_0\to \infty$, correspond to the free chain, and to the particles perfectly ordered at the bottom of the substrate potential, respectively.
In fig.~\ref{ord:fig}, we plot $\tilde a/b$, as given in (\ref{ta:def}), as a function of $a/b$, for different values of the substrate potential amplitude $V_0$ and the temperature scale. The plateaux that one observes in that figure, at $a/b =l (N-1)/N$, with $l$ an integer number, correspond to configurations where all the particles sit (on average) at the minima of the lattice potential, i.e. C-configurations.
Notice that each plateau is symmetrically  centered around  an integer value of $a/b$.
Starting from the center of one of these plateaux, as one tries to change $a/b$ from the corresponding integer value, the systems remains in the C-configurations as it costs energy to create defects (discommensurations) in the lattice. Continuing changing (increasing or decreasing) $a/b$ the strain energy increases and a transition to an IC-phase occurs. 

As expected, the width of the plateau increases and the slope of the non-constant part of the curves become sharper as i) $V_0$ is increased (see fig.~\ref{ord:fig}-(a)), ii) the temperature scale is decreased (fig.~\ref{ord:fig}-(b)). We will refer to a configuration as maximally incommensurate, when $\tilde a/b$ takes a value which is equidistant between two successive plateaux, see vertical arrows in fig.~\ref{ord:fig}-(a).
It is anticipated that the interesting motor effect in the limit of large $N$ will occur in that regime.

The shape of the curves depicted in fig.~\ref{ord:fig} is referred to as an incomplete devil’s staircase \cite{Chaikin}.
The phenomenology of the C-IC phase transition is much richer that what described here.
However a full account of the details of the C-IC transition is not the focus of this paper and the interested readers is referred to, e.g., ref.~\cite{Braun04,Chaikin}.
Furthermore, we emphasize again that the theory of the C-IC transition in the FK model (and thus the terminology C or IC phase) was derived in the limit of vanishing temperature ~\cite{Braun04, Chaikin}, while the order parameter plotted in fig.~\ref{ord:fig} is an average property of the system subject to thermal noise.

We now consider the case $\Delta T\neq 0$, and the effect of the different parameters on the system dynamic behaviour.
First we notice that the shape of the order parameter $\tilde a/b$ as a function of $a/b$ is almost unchanged for $\Delta T\neq 0$, at least for the range of parameters considered here, see fig.~\ref{ord:fig}-(c).

\begin{figure}[h]
\center
\psfrag{ft=0}[cr][cr][.8]{$\phi_T=0$}
\psfrag{ft=pi4}[cr][cr][.8]{$\phi_T=\pi/4$}
\psfrag{ft=pi2}[cr][cr][.8]{$\phi_T=\pi/2$}
\psfrag{a/b}[ct][ct][1.]{$a/b$}
\psfrag{a}[bb][bb][1.]{$a$}
\psfrag{v}[ct][ct][1.]{$\bar v$ }
\psfrag{d1}[ct][ct][.8]{$\kappa=1$}
\psfrag{d2}[ct][ct][.8]{$\kappa=2$}
\psfrag{d3}[ct][ct][.8]{$\kappa=3$}
\includegraphics[width=8cm]{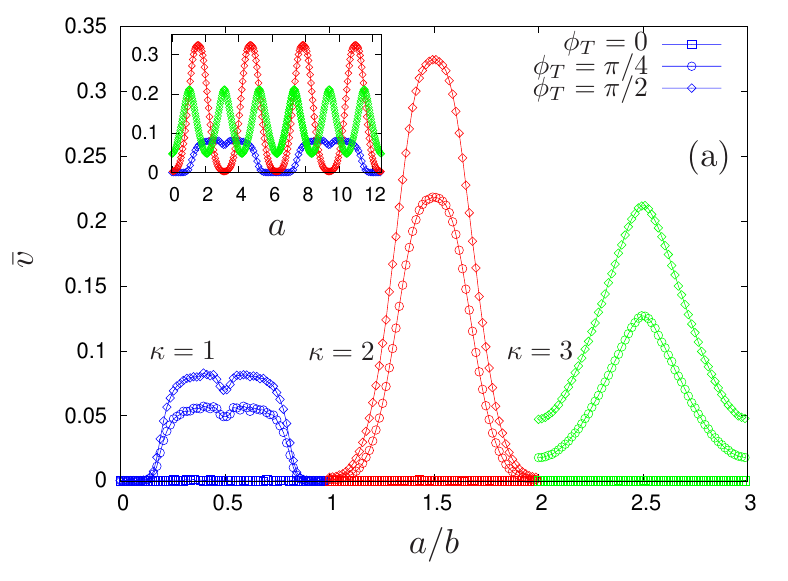}
\includegraphics[width=8cm]{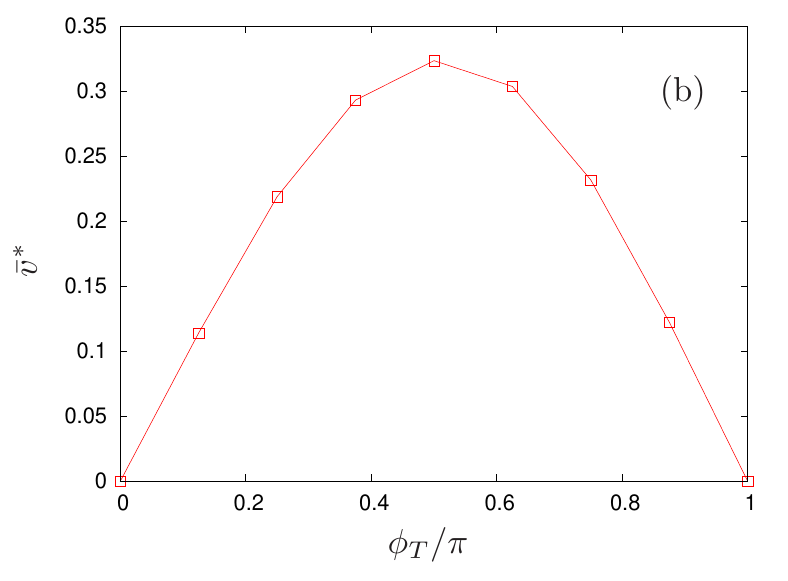}
\caption{(a): Steady state velocity $\bar v$ as a function of  the rescaled lattice spacing $a/b$  for the out-of-equilibrium FK model with $N=20$ in the temperature profile (\ref{eqT}), for different values of the phase $\phi_T$ and different values of $\kappa=\kappa_T$. The two sets of periodic curves with $\kappa=2,\,3$ are shifted for clarity. The periodicity of the velocity as a function of $a$ (with period  $2 \pi/\kappa$) is emphasized in the inset. In general the velocity profile vanishes when $\phi_T$ is a multiple integer of $\pi$ (see eq.~(\ref{eqT}) and panel (b)). (b): Maximal steady state velocity of the FK model ($\bar v^*=\max_a \bar v$) as a function of  $\phi_T$,  with $N=20$, $\kappa=\kappa_T=2$,  $\bT=1.75$, $\Delta T=1.5$, $K=V_0=5$. }
\label{fk_Tx:multipsi}
\end{figure}

In order to study the system transport dynamics, we introduce the center-of-mass steady state velocity, defined as 
\begin{equation}
\bar v=\sum_i \average{\dot x_i}/N.
\label{avv:eq}
\end{equation} 
For each set of parameters, we perform the numerical integration of eq.~(\ref{Kram:eq}) for a sufficient long time before starting to sample the velocities in eq.~(\ref{avv:eq}), so as to avoid finite time effects and be sure that the system has reached the steady state.
Figure~\ref{fk_Tx:multipsi}-(a) shows the steady state mean velocity (\ref{avv:eq})
as a function of the rescaled intra-atom equilibrium distance $a/b$ for different values of $\phi_T$ and $\kappa=2 \pi/b$.
 We notice that the velocity profile is periodic in $a$ (with period $b$), and periodic in $\phi_T$ (with period $2 \pi$),  as one would expect by inspection of eqs.~(\ref{U:def}) and (\ref{eqT}).
Furthermore when $\phi_T=n \pi$, the mean velocity is  identically zero.
One can understand this result in terms of the following symmetry argument.
Let us first consider a reflection of the coordinates about the origin and then a translation $\Delta$, and let us then require that both equalities $V(-x)=V(x+\Delta)$ and $T(-x)=T(x+\Delta)$ hold for any value of $x$. Inspection of  eq. (\ref{ext:pot}) and (\ref{eqT})  indicates that  $V(x)$ and $T(x)$ are jointly invariant under the above transformation when $\phi_T =n \pi$ and $\Delta=0$, with $n$ an integer. For any other choice of $\phi_T$ the joint spatial symmetry for $V(x)$ and $T(x)$ is broken  and the motor effect arises. 
Furthermore, such a symmetry is "maximally" broken when $\phi_T=n\pi/2$: for such value of $\phi_T$ one actually finds the larger velocity profile, see fig.~\ref{fk_Tx:multipsi}.
This result is confirmed if one fixes all the other parameters, and evaluates the maximal velocity with respect to $a$, $\bar v^*=\max_a \bar v$. In see fig.~\ref{fk_Tx:multipsi}-(b) we plot $\bar v^*$ as a function of $\phi_T$ for a given choice of the other parameters. We see that the maximum of $\bar v^*$ is indeed achieved at  $\phi_T=\pi/2$.

In non-autonomous Brownian motors driven by external mechanical forces, such as  tilting, rocked or pulsating ratchets, the substrate potential $V(x)$ alone must break the above spatial symmetry for the motor effect to appear \cite{Reimann02}. 
We finally notice that the interaction energy in eq.~(\ref{U:def}) is unaffected by either a reflection and a translation of all the coordinates. 

\begin{figure}[h]
\center
\includegraphics[width=8cm]{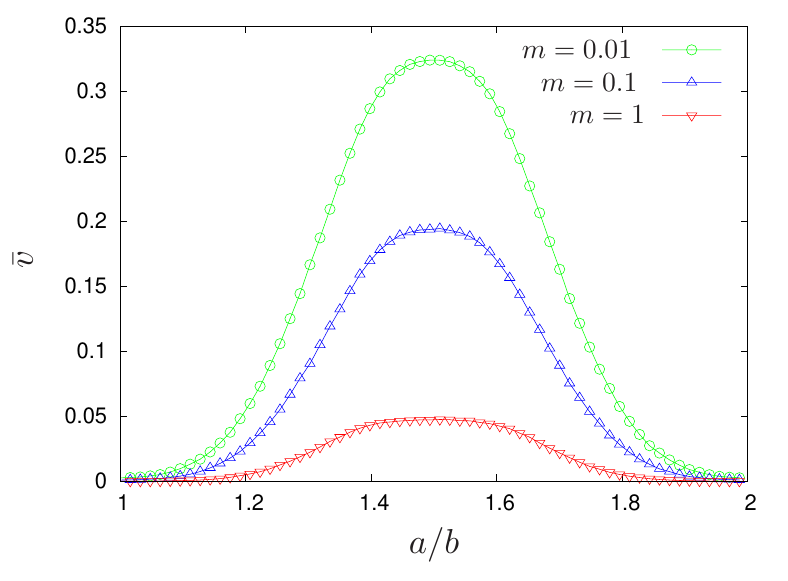}
\caption{Steady state velocity of the FK model   as a function  of $a/b$ and for different values of the mass $m$, with $N=20$, $\kappa=\kappa_T=2$, $K=V_0=5$, $\bT=1.75$, $\Delta T=1.5$, $\phi_T=\pi/2$.}
\label{multi:mass:fig}
\end{figure}

The BL model is known to work in the overdamped regime (small $m/\gamma$) while the particle current vanishes in the underdamped regime. Indeed for the motor effect to arise, the particle must equilibrate with the local environment before entering a neighbouring zone at different temperature, the typical equilibration time being given by $m/\gamma$ \cite{Blanter98}. 
The out-of-equilibrium FK model exhibits the same behaviour, as suggested by inspection of fig.~\ref{multi:mass:fig}, where the steady state velocity is plotted as a function of $a/b$ for increasing values of $m$. One clearly sees that the velocity profile goes to zero as $m$ increases.

In comparison with the BL model,  there is an additional parameter determining the dynamical behaviour of the FK model: further inspection of fig.~\ref{fk_Tx:multipsi}-(a) suggests that when $a$ is an integer multiple of $b$ (remember $\kappa=2 \pi/b$) the mean velocity is minimal.
This must be compared with the behaviour in fig.~\ref{ord:fig}-(a): when $a/b$ takes (or is close to) an  integer value, the system is in the C-phase, while for $a/b$ half integer the system is in the maximally IC-phase. This is the first important result of this paper: the motor effect is rather weak in the C-phase (it actually vanishes in the C-phase for $\kappa$ in the small-to-intermediate regime), while the  velocity is maximal in the  maximally IC-phase.



\begin{figure}[h]
\center
\psfrag{a/b}[ct][ct][1.]{$a/b$}
\psfrag{v}[ct][ct][1.]{$\bar v$ }
\psfrag{d1}[cc][cc][.7]{$\kappa=1$}
\psfrag{d1.5}[cc][cc][.7]{$\kappa=1.5$}
\psfrag{d2}[cc][cc][.7]{$\kappa=2$}
\psfrag{d2.5}[cc][cc][.7]{$\kappa=2.5$}
\psfrag{d1dt0.75}[cc][cc][.7]{$\kappa_T=0.75$}
\psfrag{d1dt1}[cc][cc][.7]{$\kappa_T=1$} 
\psfrag{d1dt2}[cc][cc][.7]{$\kappa_T=2$}
\psfrag{d1.5dt1}[cc][cc][.7]{$\kappa_T=1$}
\psfrag{d1.5dt1.5}[cc][cc][.7]{$\kappa_T=1.5$}
\psfrag{d1.5dt3}[cc][cc][.7]{$\kappa_T=3$}
\psfrag{d2dt1.5}[cc][cc][.7]{$\kappa_T=1.5$}
\psfrag{d2dt2}[cc][cc][.7]{$\kappa_T=2$}
\psfrag{d2dt2.5}[cc][cc][.7]{$\kappa_T=2.5$ }
\psfrag{d2.5dt2.25}[cc][cc][.7]{$\kappa_T=2.25$}
\psfrag{d2.5dt2.5}[cc][cc][.7]{$\kappa_T=2.5$}
\psfrag{d2.5dt2.75}[cc][cc][.7]{$\kappa_T=2.75$}
\includegraphics[width=8cm]{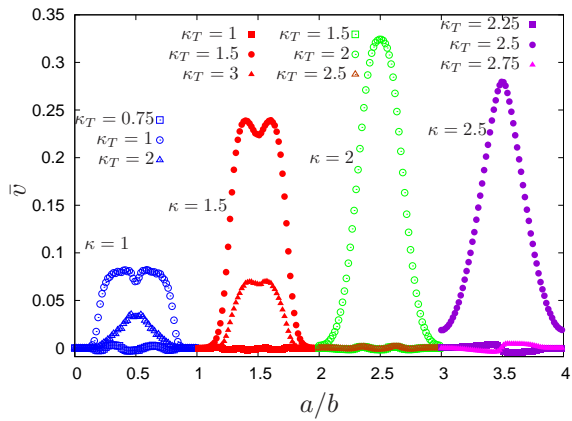}
\caption{Steady state velocity as a function of the lattice spacing $a/b$ for different values of $\kappa=2 \pi/b$ and of $\kappa_T$, with $N=20$, $\bT=1.75$, $\Delta T=1.5$, $V_0=K=5$, $\phi_T=\pi/2$. We notice that, at fixed $\kappa$ the maximum velocity profile is achieved for $\kappa_T=\kappa$. The different sets of curves are shifted for clarity.  }
\label{fk_Tx:fig}
\end{figure}

While in the equilibrium FK model the behaviour of the system is determined by the competition between 2 length scales, namely $a$ and $b$, in the present out-of-equilibrium model  a third length scale appears, namely the period of the temperature profile $2 \pi/\kappa_T$. We will now investigate the effect of this parameter on the system dynamical properties.
Inspection of fig.~\ref{fk_Tx:fig} suggests the second relevant result of this paper: at fixed $a$ and $b$, the velocity profile is larger when $\kappa=\kappa_T$, i.e. the period of the substrate and of the temperature profile must be the same for the maximal motor effect to arise.

\begin{figure}[h]
\center
\psfrag{a/b}[ct][ct][1.]{$a/b$}
\psfrag{vs}[ct][ct][.8]{$\bar v^*$ }
\psfrag{kappa}[ct][ct][.8]{$\kappa$ }
\psfrag{v}[ct][ct][1.]{$\bar v$ }
\psfrag{d1.5}[ct][ct][.8]{$\kappa=1.5$ }
\psfrag{d2}[ct][ct][.8]{$\kappa=2$ }
\psfrag{d2.5}[ct][ct][.8]{$\kappa=2.5$ }
\psfrag{np1}[cl][cl][.6]{$N=1$ }
\psfrag{np2}[cl][cl][.6]{$N=2$ }
\psfrag{np20}[cl][cl][.6]{$N=20$ }
\psfrag{np50}[cl][cl][.6]{$N=50$ }
\psfrag{np100}[cl][cl][.6]{$N=100$ }
\psfrag{np20d2}[ct][ct][.6]{$N=20$ }
\psfrag{np50d2}[ct][ct][.6]{$N=50$ }
\psfrag{np100d2}[ct][ct][.6]{$N=100$ }
\psfrag{np1000}[cl][cl][.6]{$N=1000$ }
\psfrag{np20d2.5}[ct][ct][.6]{$N=20$ }
\psfrag{np50d2.5}[ct][ct][.6]{$N=50$ }
\psfrag{np100d2.5}[ct][ct][.6]{$N=100$ }
\includegraphics[width=8cm]{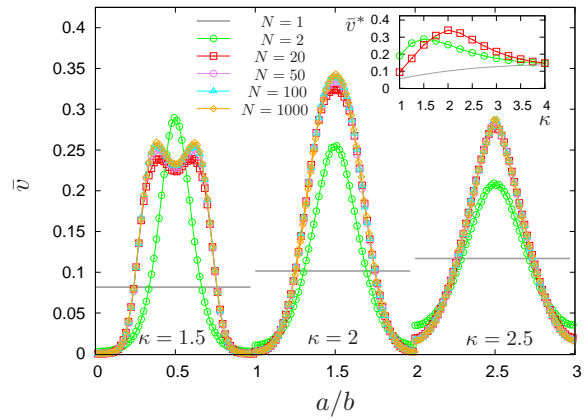}
\caption{Steady state velocity as a function of the lattice spacing $a/b$ for different values of $\kappa=\kappa_T$ and  of $N$ with $\bT=1.75$, $\Delta T=1.5$, $V_0=K=5$, $\phi_T=\pi/2$. The curves with $N=1000$ are obtained with 100 independent trajectories. We see that already at $N=20$ the curves are almost collapsed.  The different sets of curves are shifted for clarity. Inset: maximal velocity $\bar v^*$ (for $N=2$ and  20)  and velocity $\bar v$ (for $N=1$), as a function of $\kappa$.}
\label{fig:scal}
\end{figure}

We now address the next point in the analysis of the motor effect: given that the BL model already exhibits the motor effect, in which regime can the transport properties of the $N$ model outperform the $N=1$ system? 
The answer is shown in ~fig.~\ref{fig:scal}: besides the cases where the FK model is in the C-phase ($a/b\sim l$) the $N>1$ always 
outperform the BL model, in particular in the  IC-phase ($a/b$ half integer, as discussed above).
This is the third important result in this manuscript: the collective effects in the FK model enhance the motor performance with respect to the $N=1$ case, in particular in the maximally IC-phase, with the maximal degree of {\it disorder} in the chain.

Having determined the conditions under which, for a fixed $N$, the FK model exhibits the optimal velocity profile, we now analyze the scaling behaviour of the velocity profile as a function of $N$.
So far we have considered $N=20$ as the standard size for a ``large'' FK system (figs.~\ref{ord:fig}--\ref{fk_Tx:fig}), such a size having been chosen for the relatively small amount of computational time required for an extensive study of the parameter phase space.

\begin{figure}[h]
\center
\psfrag{V1}[cr][cr][.7]{$V_0=1$}
\psfrag{V5}[cr][cr][.7]{$V_0=5$}
\psfrag{V2.5}[cr][cr][.7]{$V_0=2.5$}
\psfrag{V7.5}[cr][cr][.7]{$V_0=7.5$}
\psfrag{V10}[cr][cr][.7]{$V_0=10$}
\psfrag{v}[ct][ct][1.]{$\bar v$ }
\psfrag{a/b}[ct][ct][1.]{$a/b$ }
\includegraphics[width=8cm]{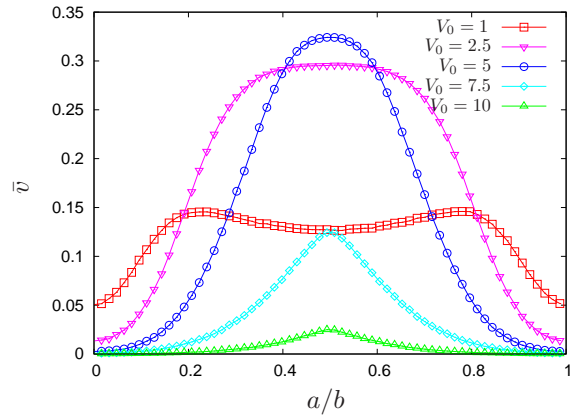}
\caption{Steady state velocity of the FK model ($N=20$) as a function of  $a/b$ for different values of the periodic potential amplitude $V_0$, with $\kappa=\kappa_T=2$,  $\bT=1.75$, $\Delta T=1.5$, $K=5$, $\phi_T=\pi/2$. }
\label{multiV0:fig}
\end{figure}

We now consider larger values of $N$. 
The results are shown in figure~\ref{fig:scal}: 
we see that already at $N=20$ the curves are almost collapsed. This justifies a posteriori the previous detailed analysis where systems with $N=20$ were taken as prototypes of the thermodynamic limit $N\to\infty$. In the inset of fig.~\ref{fig:scal} the maximal velocity of the FK model $\bar v^*=\bar v(a/b=1/2)$ is compared with the steady state of the BL model ($N=1$), which is independent of $a$. It is interesting to notice that for relatively small $\kappa$, the system with $N=2$ outperforms the manybody chain, while for moderate to large values of $\kappa$ the   manybody chain always exhibits the larger velocity profile in the IC-phase ($a/b$ half integer). A further inspection of fig.~\ref{ord:fig}-(a) suggests that for small $\kappa$ the transition between the C and the IC-phase is rather smooth, with rather narrow plateaux in the C-phase. The transition between the two phases becomes sharper as one changes $\kappa$, and given the choice of the other parameters ($\bT,\, \Delta T,\,  \, V_0,\, K$) the sharper transition is found for $\kappa=2$, which in turn gives the largest velocity profile in fig.~\ref{fig:scal}. 
One can thus conclude that in absence of a sharp C-IC transition the motor effect in the FK model is damped.
One might be then tempted to obtain an even sharper transition by increasing $V_0$: however a too large energy barrier would be an hindrance for the motion of the chain.
This effect is confirmed by inspection of fig.~\ref{multiV0:fig} where
the dependence of the steady state velocity on the periodic potential amplitude is plotted. For fixed values of the other parameters, one finds an optimal $V_0$ giving the largest velocity profile in the maximally IC-phase ($a/b$ half integer).

\begin{figure}[h]
\center
\psfrag{vs}[ct][ct][1.]{$\chi_v^*$}
\psfrag{veps}[ct][ct][1.]{$\chi_v$}
\psfrag{a/b}[ct][ct][1.]{$a/b$}
\psfrag{eps}[cl][cl][.8]{$\epsilon$}
\psfrag{e1}[cl][cl][.8]{$\epsilon=1$}
\psfrag{e2}[cl][cl][.8]{$\epsilon=1/2$}
\psfrag{e4}[cl][cl][.8]{$\epsilon=1/4$}
\psfrag{e8}[cl][cl][.8]{$\epsilon=1/8$}
\psfrag{e16}[cl][cl][.8]{$\epsilon=1/16$}
\includegraphics[width=8cm]{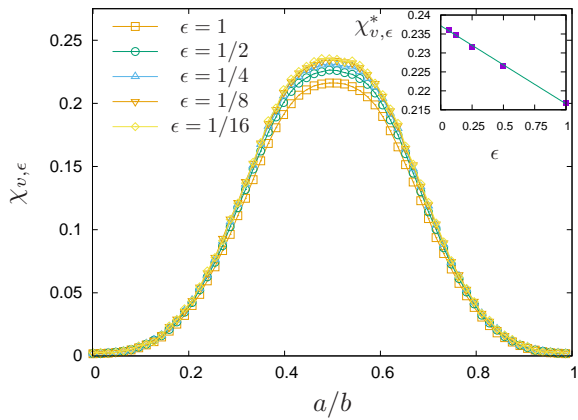}
\caption{Dynamic susceptibility (\ref{chi:def1}) as a function of $a/b$ for different values of the energy scale $\epsilon$ (see text), with $N=20$, $\tilde T=1.75$, $\widetilde{\Delta  T}=1.5$, $\tilde K=\tilde V_0=5$. Inset: maximum susceptibility $\chi_{v,\epsilon}^*=\chi_{v,\epsilon}(a/b=1/2)$ as a function of the energy scale $\epsilon$. }
\label{fig:chi}
\end{figure}

Finally, we study the scaling of the velocity profile as a function of the energy scale in the system. In particular, we use the ``dynamical susceptibility'' introduced in \cite{Sune19a} to characterize the dynamical response of a 2D model of autonomous thermal rotors at the verge of a phase transition.
For the present model we define such a quantity as 
\begin{equation}
\chi_v=\left. \derpart{\bar v}{\Delta T}\right|_{\Delta T=0}.
\label{chi:def}
\end{equation} 
One can resort to numerical differentiation, and calculate $\chi_v\simeq\bar v/\Delta T$ for decreasing values of $\Delta T$.
In particular, taking decreasing values of $\Delta T$, at fixed $K$ and $V_0$ the ratio $\bar v/\Delta T$ goes to zero as $\Delta T\to 0$ (data not shown).
We thus decrease the global energy scale in the system, by rescaling $K= \epsilon\tilde K$, $V_0=\epsilon \tilde V_0 $, $\bT =\epsilon \tilde T$,  and  $\Delta T=\epsilon \widetilde{\Delta T} $, and evaluate the steady state velocity $\bar v_\epsilon$ for decreasing values of $\epsilon$.
We then evaluate the quantity 
\begin{equation}
\chi_{v,\epsilon}=\frac{\bar v_\epsilon}{\epsilon  \widetilde{\Delta T}}.
\label{chi:def1}
\end{equation} 
The results are shown in fig.~\ref{fig:chi}: it can be seen that $\chi_{v,\epsilon}$ takes a finite value when $\epsilon \to 0$, the maximum again being achieved in the maximally IC-phase.

In the out-of-equilibrium numerical results presented in this section, the parameters $K$, $\bar T$ and $\Delta T$ have been held constant.
In appendix \ref{appb} we present some numerical results where these parameters are changed. We summarize here the main findings.
Vanishing $K$ corresponds again to the BL model. In the limit of large $K$ the chain becomes too rigid to exploit the thermal fluctuations to jump over the maxima of the substrate potential, thus suppressing the motor effect.
For a
fixed $\bT$, the motor is propelled by the temperature gradient $\Delta T$, the larger $\Delta T$, the larger the steady state velocity.
However for too large $\bT$, the system is affected by
large thermal fluctuations, which dampen the motor
effect.

\section{Thermodynamic efficiency of the out-of-equilibrium FK model}
\label{sec:eff}
\begin{figure}[h]
\center
\psfrag{f}[cr][cr][.7]{$f$}
\psfrag{f9}[cr][cr][.7]{$f=-9$}
\psfrag{f3}[cr][cr][.7]{$f=-3$}
\psfrag{f5}[cr][cr][.7]{$f=-5$}
\psfrag{f7.75}[cr][cr][.7]{$f=-7.75$}
\psfrag{a/b}[ct][ct][1.]{$a/b$}
\psfrag{pow}[ct][ct][1.]{$\Po$}
\includegraphics[width=8cm]{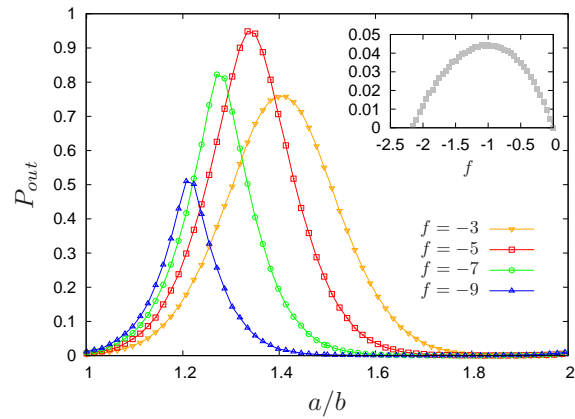}
\caption{Output power extracted from the FK model with an external load on the first particle (\ref{eq:force}) as a function  of $a/b$ and for different values of $f$, with $N=20$, $\kappa=\kappa_T=2$, $K=V_0=5$, $\bT=1.75$, $\Delta T=1.5$, $\phi_T=\pi/2$. One sees that, for this choice of parameters,  the maximum power is obtained for $f=-5$. Inset: output power of the corresponding BL model ($N=1$), as a function of the force. }
\label{fig:pow}
\end{figure}

In order to extract work, one can envisage to apply a constant force on the FK model, counteracting the motion and thus behaving as an external load.  
Assuming that the parameters are such that $\bar v>0$, work can be extracted by adding a constant negative force
to the first particle in the chain, so as eq.~(\ref{Kram:eq}) becomes
\begin{equation}
m \ddot x_i=-\partial_i U+f\delta_{i,1}-\gamma \dot x_i + \sqrt{2\gamma T(x_i)}\xi_i(t),
\label{eq:force}
\end{equation} 
with $f<0$.
One can then evaluate the extracted power $\Po= -f \bar v$, see fig.~\ref{fig:pow}.
In order to evaluate the efficiency in an engine working between two heat reservoirs, one normally needs to evaluate the heat current extracted from the hot reservoir.
Given that the FK model moves in a position dependent temperature profile $T(x)$, some care is needed.
One possible way to proceed is to consider as {\it hot} the part of substrate with $T(x)>\bT$, see eq. (\ref{eqT}).
One can thus define the current extracted by the hot surrounding environment as \cite{Fogedby12} 
\begin{equation}
\dot Q_H= \sum_i \p{p_i \dot p_i/m + (\partial_i U-f\delta_{i,0})  \dot x_i} \theta(T(x_i)-\bT),
\label{eq:QH}
\end{equation} 
where $\theta(x)$ is the Heaviside step function, and $p_i=m \dot x_i$.

\begin{figure}[h]
\center
\psfrag{f}[cr][cr][.7]{$f$}
\psfrag{f9}[cr][cr][.7]{$f=-9$}
\psfrag{f3}[cr][cr][.7]{$f=-3$}
\psfrag{f5}[cr][cr][.7]{$f=-5$}
\psfrag{f7.75}[cr][cr][.7]{$f=-7.75$}
\psfrag{eta}[ct][ct][1.]{$\eta$ }
\psfrag{a/b}[ct][ct][1.]{$a/b$ }
\includegraphics[width=8cm]{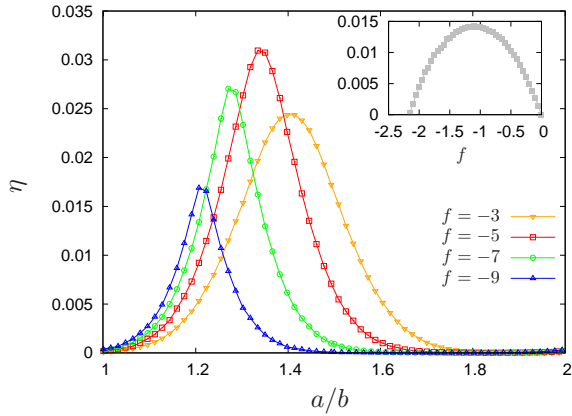}
\caption{Efficiency of the FK model with an external load on the first particle (\ref{eq:force}) as a function  of $a/b$ and for different values of $f$, with $N=20$, $\kappa=\kappa_T=2$, $K=V_0=5$, $\bT=1.75$, $\Delta T=1.5$, $\phi_T=\pi/2$, 100 time steps.  Inset: efficiency of the corresponding BL model ($N=1$), as a function of the force. }
\label{fig:eta}
\end{figure}

The efficiency is thus given by $\eta =\Po/\dot Q_H$, and can be evaluated through numerical integration of eq.~(\ref{eq:force}). A plot of $\eta$ as a function of $a/b$ and for different values of $f$ is shown in fig.~\ref{fig:eta}. Inspection of this figure suggests that the efficiency, for the set of parameters chosen here, can be as large as 3.1\%. Comparison with 
fig.~\ref{fig:pow}  indicates that this is also the value of the efficiency at maximum power (EMP).

One can compare both the output power and the efficiency of the FK model with the corresponding BL model ($N=1$), see insets in figs.~\ref{fig:pow} and \ref{fig:eta} respectively. A much larger power is extracted by the FK, bearing a larger load, and the EMP of the BL is also smaller, being $\sim$1.4\%. 

\begin{figure}[h]
\center
\psfrag{f}[cr][cr][.7]{$f$}
\psfrag{f9}[cr][cr][.7]{$f=-9$}
\psfrag{f3}[cr][cr][.7]{$f=-3$}
\psfrag{f5}[cr][cr][.7]{$f=-5$}
\psfrag{f7.75}[cr][cr][.7]{$f=-7.75$}
\psfrag{a/b}[ct][ct][1.]{$a/b$}
\psfrag{pow}[ct][ct][1.]{$\Po$}
\includegraphics[width=8cm]{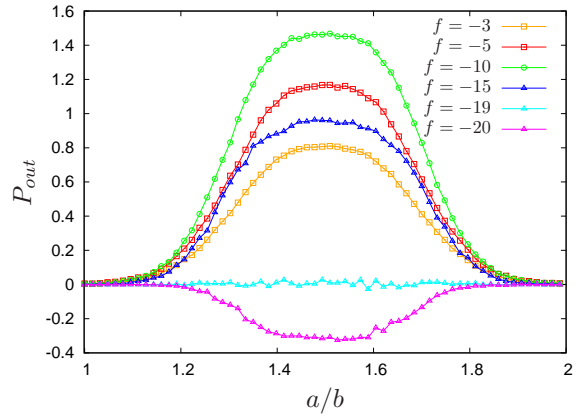}
\caption{Output power extracted from the FK model with an external load  applied on all the particles (eq.~\ref{eq:force1}), as a function  of $a/b$ and for different values of $f$, with $N=20$, $\kappa=\kappa_T=2$, $K=V_0=5$, $\bT=1.75$, $\Delta T=1.5$, $\phi_T=\pi/2$, 100 time steps. One sees that, for this choice of parameters,  the maximum power is obtained for $f=-10$. For large counteracting forces the velocity profile first vanishes, and then takes negative values, corresponding to the situation where the external force drags the chain in the negative direction. }
\label{fig:pow:all}
\end{figure}

\begin{figure}[h]
\center
\psfrag{f}[cr][cr][.7]{$f$}
\psfrag{f9}[cr][cr][.7]{$f=-9$}
\psfrag{f3}[cr][cr][.7]{$f=-3$}
\psfrag{f5}[cr][cr][.7]{$f=-5$}
\psfrag{f7.75}[cr][cr][.7]{$f=-7.75$}
\psfrag{eta}[ct][ct][1.]{$\eta$ }
\psfrag{a/b}[ct][ct][1.]{$a/b$ }
\includegraphics[width=8cm]{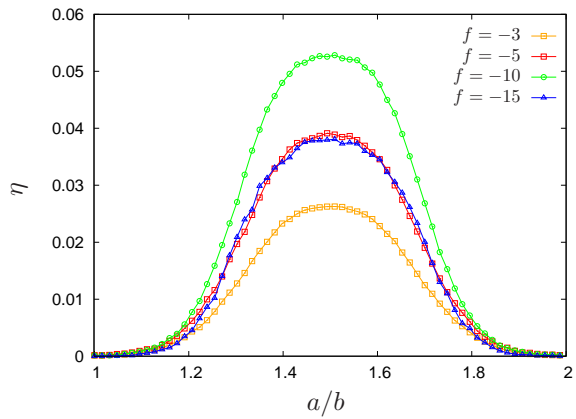}
\caption{Efficiency of the FK model with an external load on all the particles (eq.~\ref{eq:force1}), as a function  of $a/b$ and for different values of $f$, with $N=20$, $\kappa=\kappa_T=2$, $K=V_0=5$, $\bT=1.75$, $\Delta T=1.5$, $\phi_T=\pi/2$, 100 time steps.   }
\label{fig:eta:all}
\end{figure}

Another possible setup is the one where the force is evenly applied on all the particles: eqs.~(\ref{eq:force}) and (\ref{eq:QH}) thus became
\begin{equation}
m \ddot x_i=-\partial_i U+f/N-\gamma \dot x_i + \sqrt{2\gamma T(x_i)}\xi_i(t),
\label{eq:force1}
\end{equation} 
and
\begin{equation}
\dot Q_H= \sum_i \p{p_i \dot p_i/m + (\partial_i U-f/N)  \dot x_i} \theta(T(x_i)-\bT),
\label{eq:QH1}
\end{equation} 
respectively.
The corresponding results for the power and the efficiency are plotted in figs.~\ref{fig:pow:all} and \ref{fig:eta:all}, respectively. We find that, because the force is now equally distributed on all the particles, the power (fig.~\ref{fig:pow:all}) and the efficiency profiles (fig.~\ref{fig:eta:all}) are symmetric around the maximum. Furthermore, comparison with fig.~\ref{fig:pow} and ~\ref{fig:eta} indicates that the extracted power and the EMP ($\sim$5.3 \%) turn out to be larger than the ones obtained with a force applied only on the first particle as in eq.~(\ref{eq:force}). 

It is worth to note that in the present section the efficiency has been studied for a choice of parameters that maximizes the velocity $\bar v$, specifically we have taken $\kappa=\kappa_T$ and varied $a/b$. In principle other choices might lead to different values of the efficiency (possibly higher). However a full numeric exploration of the parameter space is an  impossible task given the current computational power. Thus the results presented in this section represent a bona fide estimate of the order of magnitude of the efficiency in a relevant range of the parameter space.

\section{Conclusions}
\label{sec:conc}

To conclude in this work we have studied the FK model diffusing in a position dependent temperature profile. The model exhibits a transport dynamics that depends  non-trivially on the details of the temperature profile $T(x)$,  of the substrate shape, as represented by $V(x)$ in eq.~(\ref{ext:pot}), and on  the particle-particle equilibrium distance $a$.
The model outperforms the BL model (monomer) whenever in the IC-phase, thus confirming the relevance of the collective effects in models of interacting motors.

As such, the out-of-equilibrium FK model discussed in this paper represents an extremely efficient setup to transport a large number of particles, somehow contradicting the common intuition  that a large 1D chain of particles may form a clog and thus impede mass transport.
In this respect, the model resembles the asymmetric simple exclusion process (ASEP) \cite{Derrida1993,Schutz1993},  an interacting particle system  with particles diffusing on a 1D  lattice, the term exclusion indicating that two particles cannot occupy the same site. Similarly to the FK model considered here, in the ASEP the parameters can be tuned so as to have a maximal current phase, as opposed to a high-density phase with a lower particle current due to particle queuing.

Systems of optically trapped Brownian particles, where the local temperature can be controlled by modulating the trapping laser, have been experimentally studied in \cite{Berut16,Berut16a}.
Experimentally, the FK potential (\ref{U:def}) has been used to describe coulomb crystals of ions in optical lattices \cite{Bylinskii2015,Gangloff2015,Bylinskii2016,Kiethe2018}.
In particular, in \cite{Bylinskii2015} the authors were able to control  the nanofriction at the individual-atom level. A similar setup might be used to control the local temperature. 
In \cite{Drewsen19} a possible experimental setup with a duet of sideband laser--cooled atomic ions  ($N=2$) was discussed: by keeping the ions at different temperatures the  center of mass is expected to exhibit a nonvanishing velocity.

Ion-crystal simulators represent thus a possible setup for an experimental realization of the model proposed in the present manuscript.

\appendix
\section{The BL model}
\label{appa}

The B\"uttiker-Landauer model \cite{Buttiker1987,Landauer1988} consists of  a single Brownian particle diffusing in a periodic potential $V(x)$, with a position-dependent and periodic temperature profile $T(x)$.
As discussed in the main text, for the specific
case where  $V(x)$ and $T(x)$ have the same period $L$, it has been shown that the single particle exhibits a net non-zero velocity if the effective potential $u(x)=\int_0^{x} \D y\,  V'(y)/T(y)$ is not a periodic function of $x$ \cite{Matsuo2000,Fogedby17}. 

\begin{figure}[h]
\center
\psfrag{psi0}[cr][cr][.8]{$\phi_T=0$}
\psfrag{psi4}[cr][cr][.8]{$\phi_T=\pi/4$}
\psfrag{psi2}[cr][cr][.8]{$\phi_T=\pi/2$}
\psfrag{v}[ct][ct][1.]{$\bar v$ }
\psfrag{k}[ct][ct][1.]{$\kappa$ }
\includegraphics[width=8cm]{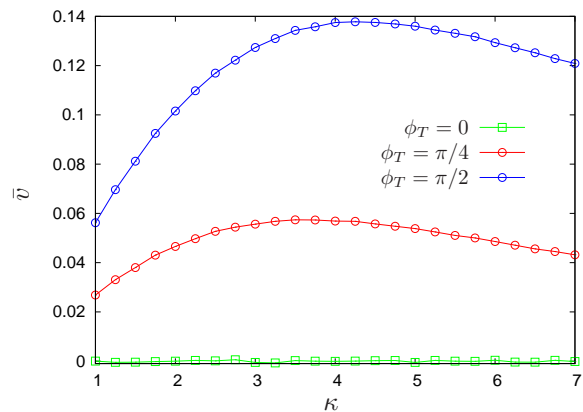}
\caption{Steady state velocity of the BL model ($N=1$) as a function of the substrate wavenumber $\kappa$ for different values of $\phi_T$,  with $\kappa_T=\kappa$, $\bT=1.75$, $\Delta T=1.5$, $V_0=5$.}
\label{vnp1:fig}
\end{figure}

As an illustrative example of non-periodic effective potential in the BL model, one can consider the case, $V(x)=-V_0 \cos\kappa x$, and $T(x)=\overline T +\Delta T \cos(\kappa_T x+\phi_T)$, with $\Delta T<\overline T$, and  $\kappa = l \kappa_T$  where $l$ is a positive integer. This is thus a more general case than that considered in \cite{Matsuo2000,Fogedby17} where $l=1$. $V'(x)/T(x)$ is thus periodic with period $L=2 \pi/\kappa_T$.  By   expanding $V'(x)/T(x)$ in series of $\Delta T $, one finds that the term of order $n$ in the series is proportional to  $ (\Delta T \cos(\kappa_T x))^n \sin \kappa x$. The corresponding term in the series of $u(L)=\int_0^{L} \D y\,  V'(y)/T(y)$ is 
$\delta_{l,n} (\Delta T)^n \sin n \phi_T$. Thus, the only non-zero term is proportional to $ \sin l \phi_T$, and we conclude that $u(x)$ is periodic ($u(0)=u(L)=0$) only if $\phi_T$ is a multiple integer of $\pi/l$.
In figure~\ref{vnp1:fig}, we plot the velocity of the BL model as a function of $\kappa$, for different values of $\phi_T$ (here we take $\kappa_T=\kappa$), confirming that the steady state velocity vanishes when $\phi_T$ is a multiple integer of $\pi/l$.

On the other hand, if the potential $V(x)$ and the temperature profile $T(x)$ exhibit the symmetry discussed in the main text, 
i.e. that it is possible to find a $\Delta$ such that the equalities  $V(-x)=V(x+\Delta)$ and $T(-x)=T(x+\Delta)$ hold for any value of $x$, one can show that $u(0)=u(L)$.
Indeed, in this case one has
\begin{eqnarray}
u(L)&=&\int_0^{L} \D y\,  V'(y)/T(y)=\int_{-L}^{0} \D y\,  V'(-y)/T(-y)\nonumber\\
&=&-\int^\Delta_{\Delta-L} \D y\,  V'(y)/T(y).
\label{eq:appa}
\end{eqnarray} 
By expressing  the periodic function $V'(y)/T(y)$ as a Fourier series,  
\begin{equation}
V'(y)/T(y)=h_0+ \sum_{q \neq 0} \E^{i q  y}  h_q,
\end{equation} 
with $q=2\pi l/L$, 
one sees that eq.~(\ref{eq:appa}) implies $h_0=0$, and thus  $u(0)=u(L)$.

\section{Additional results on the FK model}
\label{appb}

\begin{figure}[h]
\center
\includegraphics[width=8cm]{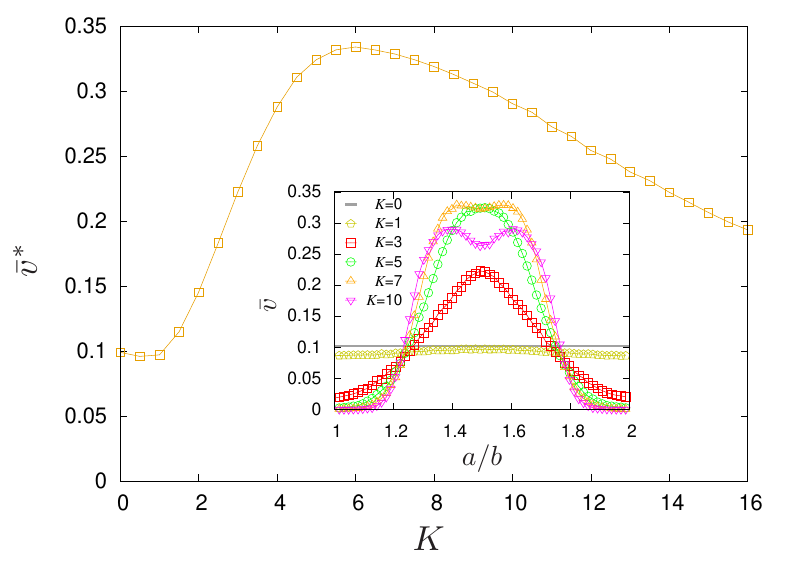}
\caption{Maximal steady state velocity $\bar v^*=\max_a \bar v$ as a function of the harmonic interaction strength $K$, for  $N=20$, $\kappa=\kappa_T=2$, $\bT=1.75$, $\Delta T=1.5$, $V_0=5$, $\phi_T=\pi/2$, 100 time steps.  Inset:  velocity profile $\bar v$  as a function of the rescaled lattice spacing $a/b$ for different values of $K$. }
\label{multiK:fig}
\end{figure}

\begin{figure}[h]
\center
\includegraphics[width=8cm]{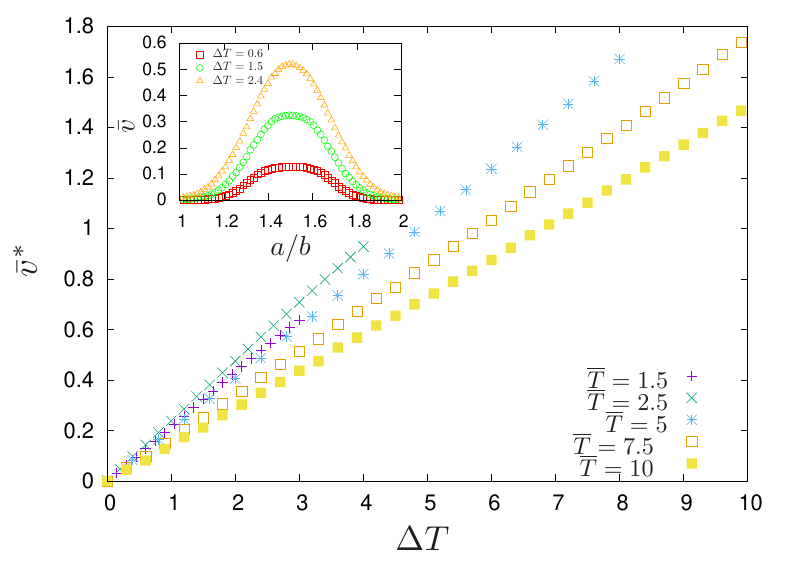}
\caption{Maximal steady state velocity $\bar v^*=\max_a \bar v$ as a function of the temperature gradient $\Delta T$, for different values of the mean velocity $\bT$, and for  $N=20$, $\kappa=\kappa_T=2$,  $V_0=K=5$, $\phi_T=\pi/2$, 100 time steps.  Inset:  velocity profile $\bar v$  as a function of the rescaled lattice spacing $a/b$ for different values of $\Delta T$, with $\bT=1.75$. }
\label{multiDT:fig}
\end{figure}

The out-of-equilibrium FK model presented in this paper is characterized by  several energy scales, namely the periodic potential
amplitude $V_0$, the mean temperature $\bT$ and the temperature gradient $\Delta T$, as well as the  harmonic potential stiffness $K$ (the last parameter having dimensions energy/length$^2$).
In the main text we fixed the values of the last three parameters, and studied the steady state velocity for different $V_0$, see fig.~\ref{multiV0:fig}. In the following we show some results obtained by varying the remaining three parameters. \\
Focusing first on $K$, one can expect that for $K=0$ the BL result is recovered ($N$ non interacting particles), while for large $K$ the chain becomes stiffer and stiffer, so as the motor effect is suppressed. This is confirmed by inspection of fig.~(\ref{multiK:fig}), where we plot the steady state velocity as a function of $a/b$ for different values of $K$, and find that the optimal working regime occurs at intermediate values of $K$.\\
The interplay between $\bT$ and $\Delta T$ is more subtle: for a fixed $\bT$, the motor is propelled by the temperature gradient  $\Delta T$, the larger  $\Delta T$, the larger the steady state velocity. However for too large $\bT$, the system is affected by large thermal fluctuations, which will suppress the motor effect: the FK chain will randomly fluctuate over the substrate potential without a preferred direction. Thus the optimal working regime for a given value of $\Delta T$ is found at intermediate values of $\bT$, as confirmed by inspection of fig.~\ref{multiDT:fig}. Notice that for a given value of $\bT$, the temperature gradient is bounded by the condition $\bT -\Delta T/2>0$, see eq.~(\ref{eqT}).

\newpage
\bibliography{bibliography}

\begin{thebibliography}{39}
\expandafter\ifx\csname natexlab\endcsname\relax\def\natexlab#1{#1}\fi
\expandafter\ifx\csname bibnamefont\endcsname\relax
  \def\bibnamefont#1{#1}\fi
\expandafter\ifx\csname bibfnamefont\endcsname\relax
  \def\bibfnamefont#1{#1}\fi
\expandafter\ifx\csname citenamefont\endcsname\relax
  \def\citenamefont#1{#1}\fi
\expandafter\ifx\csname url\endcsname\relax
  \def\url#1{\texttt{#1}}\fi
\expandafter\ifx\csname urlprefix\endcsname\relax\def\urlprefix{URL }\fi
\providecommand{\bibinfo}[2]{#2}
\providecommand{\eprint}[2][]{\url{#2}}

\bibitem[{\citenamefont{Frenkel and Kontorova}(1938{\natexlab{a}})}]{FK38}
\bibinfo{author}{\bibfnamefont{Y.}~\bibnamefont{Frenkel}} \bibnamefont{and}
  \bibinfo{author}{\bibfnamefont{T.}~\bibnamefont{Kontorova}},
  \bibinfo{journal}{Zh. Eksp. Teor. Fiz.} \textbf{\bibinfo{volume}{8}},
  \bibinfo{pages}{89} (\bibinfo{year}{1938}{\natexlab{a}}).

\bibitem[{\citenamefont{Frenkel and Kontorova}(1938{\natexlab{b}})}]{FK38a}
\bibinfo{author}{\bibfnamefont{Y.}~\bibnamefont{Frenkel}} \bibnamefont{and}
  \bibinfo{author}{\bibfnamefont{T.}~\bibnamefont{Kontorova}},
  \bibinfo{journal}{Zh. Eksp. Teor. Fiz.} \textbf{\bibinfo{volume}{8}},
  \bibinfo{pages}{1340} (\bibinfo{year}{1938}{\natexlab{b}}).

\bibitem[{\citenamefont{Frank and van~der Merwe}(1949)}]{Frank49}
\bibinfo{author}{\bibfnamefont{F.~C.} \bibnamefont{Frank}} \bibnamefont{and}
  \bibinfo{author}{\bibfnamefont{J.~H.} \bibnamefont{van~der Merwe}},
  \bibinfo{journal}{Proceedings of the Royal Society of London. Series A.
  Mathematical and Physical Sciences} \textbf{\bibinfo{volume}{198}},
  \bibinfo{pages}{205} (\bibinfo{year}{1949}),
  \eprint{https://royalsocietypublishing.org/doi/pdf/10.1098/rspa.1949.0095},
  \urlprefix\url{https://royalsocietypublishing.org/doi/abs/10.1098/rspa.1949.0095}.

\bibitem[{\citenamefont{Braun and Kivshar}(2004)}]{Braun04}
\bibinfo{author}{\bibfnamefont{O.~M.} \bibnamefont{Braun}} \bibnamefont{and}
  \bibinfo{author}{\bibfnamefont{Y.~S.} \bibnamefont{Kivshar}},
  \emph{\bibinfo{title}{The Frenkel-Kontorova Model Concepts, Methods, and
  Applications}}, Theoretical and Mathematical Physics
  (\bibinfo{publisher}{Springer Berlin Heidelberg}, \bibinfo{address}{Berlin,
  Heidelberg}, \bibinfo{year}{2004}), \bibinfo{edition}{1st} ed., ISBN
  \bibinfo{isbn}{3-662-10331-1}.

\bibitem[{\citenamefont{van~den Ende et~al.}(2012)\citenamefont{van~den Ende,
  de~Wijn, and Fasolino}}]{van_den_Ende_2012}
\bibinfo{author}{\bibfnamefont{J.~A.} \bibnamefont{van~den Ende}},
  \bibinfo{author}{\bibfnamefont{A.~S.} \bibnamefont{de~Wijn}},
  \bibnamefont{and} \bibinfo{author}{\bibfnamefont{A.}~\bibnamefont{Fasolino}},
  \bibinfo{journal}{Journal of Physics: Condensed Matter}
  \textbf{\bibinfo{volume}{24}}, \bibinfo{pages}{445009}
  (\bibinfo{year}{2012}),
  \urlprefix\url{https://doi.org/10.1088%2F0953-8984%2F24%2F44%2F445009}.

\bibitem[{\citenamefont{Norell et~al.}(2016)\citenamefont{Norell, Fasolino, and
  de~Wijn}}]{Norell16}
\bibinfo{author}{\bibfnamefont{J.}~\bibnamefont{Norell}},
  \bibinfo{author}{\bibfnamefont{A.}~\bibnamefont{Fasolino}}, \bibnamefont{and}
  \bibinfo{author}{\bibfnamefont{A.~S.} \bibnamefont{de~Wijn}},
  \bibinfo{journal}{Phys. Rev. E} \textbf{\bibinfo{volume}{94}},
  \bibinfo{pages}{023001} (\bibinfo{year}{2016}),
  \urlprefix\url{https://link.aps.org/doi/10.1103/PhysRevE.94.023001}.

\bibitem[{\citenamefont{Hu et~al.}(1998)\citenamefont{Hu, Li, and Zhao}}]{Hu98}
\bibinfo{author}{\bibfnamefont{B.}~\bibnamefont{Hu}},
  \bibinfo{author}{\bibfnamefont{B.}~\bibnamefont{Li}}, \bibnamefont{and}
  \bibinfo{author}{\bibfnamefont{H.}~\bibnamefont{Zhao}},
  \bibinfo{journal}{Phys. Rev. E} \textbf{\bibinfo{volume}{57}},
  \bibinfo{pages}{2992} (\bibinfo{year}{1998}),
  \urlprefix\url{https://link.aps.org/doi/10.1103/PhysRevE.57.2992}.

\bibitem[{\citenamefont{Lepri et~al.}(2003)\citenamefont{Lepri, Livi, and
  Politi}}]{Lepri03}
\bibinfo{author}{\bibfnamefont{S.}~\bibnamefont{Lepri}},
  \bibinfo{author}{\bibfnamefont{R.}~\bibnamefont{Livi}}, \bibnamefont{and}
  \bibinfo{author}{\bibfnamefont{A.}~\bibnamefont{Politi}},
  \bibinfo{journal}{Physics Reports} \textbf{\bibinfo{volume}{377}},
  \bibinfo{pages}{1 } (\bibinfo{year}{2003}), ISSN \bibinfo{issn}{0370-1573},
  \urlprefix\url{http://www.sciencedirect.com/science/article/pii/S0370157302005586}.

\bibitem[{\citenamefont{Bylinskii et~al.}(2015)\citenamefont{Bylinskii,
  Gangloff, and Vuleti{\'c}}}]{Bylinskii2015}
\bibinfo{author}{\bibfnamefont{A.}~\bibnamefont{Bylinskii}},
  \bibinfo{author}{\bibfnamefont{D.}~\bibnamefont{Gangloff}}, \bibnamefont{and}
  \bibinfo{author}{\bibfnamefont{V.}~\bibnamefont{Vuleti{\'c}}},
  \bibinfo{journal}{Science} \textbf{\bibinfo{volume}{348}},
  \bibinfo{pages}{1115} (\bibinfo{year}{2015}), ISSN \bibinfo{issn}{0036-8075},
  \eprint{https://science.sciencemag.org/content/348/6239/1115.full.pdf},
  \urlprefix\url{https://science.sciencemag.org/content/348/6239/1115}.

\bibitem[{\citenamefont{Gangloff et~al.}(2015)\citenamefont{Gangloff,
  Bylinskii, Counts, Jhe, and Vuleti{\'{c}}}}]{Gangloff2015}
\bibinfo{author}{\bibfnamefont{D.}~\bibnamefont{Gangloff}},
  \bibinfo{author}{\bibfnamefont{A.}~\bibnamefont{Bylinskii}},
  \bibinfo{author}{\bibfnamefont{I.}~\bibnamefont{Counts}},
  \bibinfo{author}{\bibfnamefont{W.}~\bibnamefont{Jhe}}, \bibnamefont{and}
  \bibinfo{author}{\bibfnamefont{V.}~\bibnamefont{Vuleti{\'{c}}}},
  \bibinfo{journal}{Nature Physics} \textbf{\bibinfo{volume}{11}},
  \bibinfo{pages}{915} (\bibinfo{year}{2015}), ISSN \bibinfo{issn}{1745-2481},
  \urlprefix\url{https://doi.org/10.1038/nphys3459}.

\bibitem[{\citenamefont{Bylinskii et~al.}(2016)\citenamefont{Bylinskii,
  Gangloff, Counts, and Vuleti{\'{c}}}}]{Bylinskii2016}
\bibinfo{author}{\bibfnamefont{A.}~\bibnamefont{Bylinskii}},
  \bibinfo{author}{\bibfnamefont{D.}~\bibnamefont{Gangloff}},
  \bibinfo{author}{\bibfnamefont{I.}~\bibnamefont{Counts}}, \bibnamefont{and}
  \bibinfo{author}{\bibfnamefont{V.}~\bibnamefont{Vuleti{\'{c}}}},
  \bibinfo{journal}{Nature Materials} \textbf{\bibinfo{volume}{15}},
  \bibinfo{pages}{717} (\bibinfo{year}{2016}), ISSN \bibinfo{issn}{1476-4660},
  \urlprefix\url{https://doi.org/10.1038/nmat4601}.

\bibitem[{\citenamefont{Kiethe et~al.}(2018)\citenamefont{Kiethe, Nigmatullin,
  Schmirander, Kalincev, and Mehlstäubler}}]{Kiethe2018}
\bibinfo{author}{\bibfnamefont{J.}~\bibnamefont{Kiethe}},
  \bibinfo{author}{\bibfnamefont{R.}~\bibnamefont{Nigmatullin}},
  \bibinfo{author}{\bibfnamefont{T.}~\bibnamefont{Schmirander}},
  \bibinfo{author}{\bibfnamefont{D.}~\bibnamefont{Kalincev}}, \bibnamefont{and}
  \bibinfo{author}{\bibfnamefont{T.~E.} \bibnamefont{Mehlstäubler}},
  \bibinfo{journal}{New Journal of Physics} \textbf{\bibinfo{volume}{20}},
  \bibinfo{pages}{123017} (\bibinfo{year}{2018}),
  \urlprefix\url{https://doi.org/10.1088%2F1367-2630%2Faaf3d5}.

\bibitem[{\citenamefont{Mu{\~n}oz et~al.}(2020)\citenamefont{Mu{\~n}oz, Sawant,
  Maffei, Wanx, and Barontini}}]{Munoz2020}
\bibinfo{author}{\bibfnamefont{J.~M.} \bibnamefont{Mu{\~n}oz}},
  \bibinfo{author}{\bibfnamefont{R.}~\bibnamefont{Sawant}},
  \bibinfo{author}{\bibfnamefont{A.}~\bibnamefont{Maffei}},
  \bibinfo{author}{\bibfnamefont{X.}~\bibnamefont{Wanx}}, \bibnamefont{and}
  \bibinfo{author}{\bibfnamefont{G.}~\bibnamefont{Barontini}},
  \bibinfo{journal}{arXiv:2006.16304}  (\bibinfo{year}{2020}).

\bibitem[{\citenamefont{Chaikin and Lubensky}(1995)}]{Chaikin}
\bibinfo{author}{\bibfnamefont{P.~M.} \bibnamefont{Chaikin}} \bibnamefont{and}
  \bibinfo{author}{\bibfnamefont{T.~C.} \bibnamefont{Lubensky}},
  \emph{\bibinfo{title}{Principles of Condensed Matter Physics}}
  (\bibinfo{publisher}{Cambridge University Press},
  \bibinfo{address}{Cambridge}, \bibinfo{year}{1995}).

\bibitem[{\citenamefont{Campisi and Fazio}(2016)}]{Campisi2016}
\bibinfo{author}{\bibfnamefont{M.}~\bibnamefont{Campisi}} \bibnamefont{and}
  \bibinfo{author}{\bibfnamefont{R.}~\bibnamefont{Fazio}},
  \bibinfo{journal}{Nat. Comm.} \textbf{\bibinfo{volume}{7}},
  \bibinfo{pages}{11895 EP } (\bibinfo{year}{2016}), \bibinfo{note}{article},
  \urlprefix\url{http://dx.doi.org/10.1038/ncomms11895}.

\bibitem[{\citenamefont{Golubeva and Imparato}(2012)}]{Golubeva2012a}
\bibinfo{author}{\bibfnamefont{N.}~\bibnamefont{Golubeva}} \bibnamefont{and}
  \bibinfo{author}{\bibfnamefont{A.}~\bibnamefont{Imparato}},
  \bibinfo{journal}{Phys. Rev. Lett.} \textbf{\bibinfo{volume}{109}},
  \bibinfo{pages}{190602} (\bibinfo{year}{2012}).

\bibitem[{\citenamefont{Golubeva and Imparato}(2013)}]{Golubeva2013}
\bibinfo{author}{\bibfnamefont{N.}~\bibnamefont{Golubeva}} \bibnamefont{and}
  \bibinfo{author}{\bibfnamefont{A.}~\bibnamefont{Imparato}},
  \bibinfo{journal}{Phys. Rev. E} \textbf{\bibinfo{volume}{88}},
  \bibinfo{pages}{012114} (\bibinfo{year}{2013}).

\bibitem[{\citenamefont{Golubeva and Imparato}(2014)}]{Golubeva2014}
\bibinfo{author}{\bibfnamefont{N.}~\bibnamefont{Golubeva}} \bibnamefont{and}
  \bibinfo{author}{\bibfnamefont{A.}~\bibnamefont{Imparato}},
  \bibinfo{journal}{Phys. Rev. E} \textbf{\bibinfo{volume}{89}},
  \bibinfo{pages}{062118} (\bibinfo{year}{2014}),
  \urlprefix\url{http://link.aps.org/doi/10.1103/PhysRevE.89.062118}.

\bibitem[{\citenamefont{Imparato}(2015)}]{Imparato15}
\bibinfo{author}{\bibfnamefont{A.}~\bibnamefont{Imparato}},
  \bibinfo{journal}{New Journal of Physics} \textbf{\bibinfo{volume}{17}},
  \bibinfo{pages}{125004} (\bibinfo{year}{2015}),
  \urlprefix\url{http://stacks.iop.org/1367-2630/17/i=12/a=125004}.

\bibitem[{\citenamefont{Herpich et~al.}(2018)\citenamefont{Herpich, Thingna,
  and Esposito}}]{Herpich18}
\bibinfo{author}{\bibfnamefont{T.}~\bibnamefont{Herpich}},
  \bibinfo{author}{\bibfnamefont{J.}~\bibnamefont{Thingna}}, \bibnamefont{and}
  \bibinfo{author}{\bibfnamefont{M.}~\bibnamefont{Esposito}},
  \bibinfo{journal}{Phys. Rev. X} \textbf{\bibinfo{volume}{8}},
  \bibinfo{pages}{031056} (\bibinfo{year}{2018}),
  \urlprefix\url{https://link.aps.org/doi/10.1103/PhysRevX.8.031056}.

\bibitem[{\citenamefont{Herpich and Esposito}(2019)}]{Herpich18a}
\bibinfo{author}{\bibfnamefont{T.}~\bibnamefont{Herpich}} \bibnamefont{and}
  \bibinfo{author}{\bibfnamefont{M.}~\bibnamefont{Esposito}},
  \bibinfo{journal}{Phys. Rev. E} \textbf{\bibinfo{volume}{99}},
  \bibinfo{pages}{022135} (\bibinfo{year}{2019}),
  \urlprefix\url{https://link.aps.org/doi/10.1103/PhysRevE.99.022135}.

\bibitem[{\citenamefont{Su{\~{n}}{\'e} and Imparato}(2019)}]{SuneImparato19}
\bibinfo{author}{\bibfnamefont{M.}~\bibnamefont{Su{\~{n}}{\'e}}}
  \bibnamefont{and} \bibinfo{author}{\bibfnamefont{A.}~\bibnamefont{Imparato}},
  \bibinfo{journal}{Journal of Physics A: Mathematical and Theoretical}
  \textbf{\bibinfo{volume}{52}}, \bibinfo{pages}{045003}
  (\bibinfo{year}{2019}),
  \urlprefix\url{http://stacks.iop.org/1751-8121/52/i=4/a=045003}.

\bibitem[{\citenamefont{Fogedby and Imparato}(2017)}]{Fogedby17}
\bibinfo{author}{\bibfnamefont{H.~C.} \bibnamefont{Fogedby}} \bibnamefont{and}
  \bibinfo{author}{\bibfnamefont{A.}~\bibnamefont{Imparato}},
  \bibinfo{journal}{EPL (Europhysics Letters)} \textbf{\bibinfo{volume}{119}},
  \bibinfo{pages}{50007} (\bibinfo{year}{2017}),
  \urlprefix\url{http://stacks.iop.org/0295-5075/119/i=5/a=50007}.

\bibitem[{\citenamefont{Fogedby and Imparato}(2018)}]{Fogedby18}
\bibinfo{author}{\bibfnamefont{H.~C.} \bibnamefont{Fogedby}} \bibnamefont{and}
  \bibinfo{author}{\bibfnamefont{A.}~\bibnamefont{Imparato}},
  \bibinfo{journal}{EPL (Europhysics Letters)} \textbf{\bibinfo{volume}{122}},
  \bibinfo{pages}{10006} (\bibinfo{year}{2018}),
  \urlprefix\url{http://stacks.iop.org/0295-5075/122/i=1/a=10006}.

\bibitem[{\citenamefont{Su\~n\'e and Imparato}(2019)}]{Sune19a}
\bibinfo{author}{\bibfnamefont{M.}~\bibnamefont{Su\~n\'e}} \bibnamefont{and}
  \bibinfo{author}{\bibfnamefont{A.}~\bibnamefont{Imparato}},
  \bibinfo{journal}{Phys. Rev. Lett.} \textbf{\bibinfo{volume}{123}},
  \bibinfo{pages}{070601} (\bibinfo{year}{2019}),
  \urlprefix\url{https://link.aps.org/doi/10.1103/PhysRevLett.123.070601}.

\bibitem[{\citenamefont{Drewsen and Imparato}(2019)}]{Drewsen19}
\bibinfo{author}{\bibfnamefont{M.}~\bibnamefont{Drewsen}} \bibnamefont{and}
  \bibinfo{author}{\bibfnamefont{A.}~\bibnamefont{Imparato}},
  \bibinfo{journal}{Phys. Rev. E} \textbf{\bibinfo{volume}{100}},
  \bibinfo{pages}{042138} (\bibinfo{year}{2019}),
  \urlprefix\url{https://link.aps.org/doi/10.1103/PhysRevE.100.042138}.

\bibitem[{\citenamefont{B{\"u}ttiker}(1987)}]{Buttiker1987}
\bibinfo{author}{\bibfnamefont{M.}~\bibnamefont{B{\"u}ttiker}},
  \bibinfo{journal}{Zeitschrift f{\"u}r Physik B Condensed Matter}
  \textbf{\bibinfo{volume}{68}}, \bibinfo{pages}{161} (\bibinfo{year}{1987}),
  ISSN \bibinfo{issn}{1431-584X},
  \urlprefix\url{http://dx.doi.org/10.1007/BF01304221}.

\bibitem[{\citenamefont{Landauer}(1988)}]{Landauer1988}
\bibinfo{author}{\bibfnamefont{R.}~\bibnamefont{Landauer}},
  \bibinfo{journal}{Journal of Statistical Physics}
  \textbf{\bibinfo{volume}{53}}, \bibinfo{pages}{233} (\bibinfo{year}{1988}),
  ISSN \bibinfo{issn}{1572-9613},
  \urlprefix\url{http://dx.doi.org/10.1007/BF01011555}.

\bibitem[{\citenamefont{Matsuo and ichi Sasa}(2000)}]{Matsuo2000}
\bibinfo{author}{\bibfnamefont{M.}~\bibnamefont{Matsuo}} \bibnamefont{and}
  \bibinfo{author}{\bibfnamefont{S.}~\bibnamefont{ichi Sasa}},
  \bibinfo{journal}{Physica A: Statistical Mechanics and its Applications}
  \textbf{\bibinfo{volume}{276}}, \bibinfo{pages}{188 } (\bibinfo{year}{2000}),
  ISSN \bibinfo{issn}{0378-4371},
  \urlprefix\url{http://www.sciencedirect.com/science/article/pii/S0378437199003659}.

\bibitem[{\citenamefont{Der\'enyi and Astumian}(1999)}]{Derenyi99}
\bibinfo{author}{\bibfnamefont{I.}~\bibnamefont{Der\'enyi}} \bibnamefont{and}
  \bibinfo{author}{\bibfnamefont{R.~D.} \bibnamefont{Astumian}},
  \bibinfo{journal}{Phys. Rev. E} \textbf{\bibinfo{volume}{59}},
  \bibinfo{pages}{R6219} (\bibinfo{year}{1999}),
  \urlprefix\url{https://link.aps.org/doi/10.1103/PhysRevE.59.R6219}.

\bibitem[{\citenamefont{Hondou and Sekimoto}(2000)}]{Hondou2000}
\bibinfo{author}{\bibfnamefont{T.}~\bibnamefont{Hondou}} \bibnamefont{and}
  \bibinfo{author}{\bibfnamefont{K.}~\bibnamefont{Sekimoto}},
  \bibinfo{journal}{Phys. Rev. E} \textbf{\bibinfo{volume}{62}},
  \bibinfo{pages}{6021} (\bibinfo{year}{2000}),
  \urlprefix\url{http://link.aps.org/doi/10.1103/PhysRevE.62.6021}.

\bibitem[{\citenamefont{Benjamin and Kawai}(2008)}]{Benjamin08}
\bibinfo{author}{\bibfnamefont{R.}~\bibnamefont{Benjamin}} \bibnamefont{and}
  \bibinfo{author}{\bibfnamefont{R.}~\bibnamefont{Kawai}},
  \bibinfo{journal}{Phys. Rev. E} \textbf{\bibinfo{volume}{77}},
  \bibinfo{pages}{051132} (\bibinfo{year}{2008}),
  \urlprefix\url{http://link.aps.org/doi/10.1103/PhysRevE.77.051132}.

\bibitem[{\citenamefont{Reimann}(2002)}]{Reimann02}
\bibinfo{author}{\bibfnamefont{P.}~\bibnamefont{Reimann}},
  \bibinfo{journal}{Physics Reports} \textbf{\bibinfo{volume}{361}},
  \bibinfo{pages}{57 } (\bibinfo{year}{2002}), ISSN \bibinfo{issn}{0370 -
  1573},
  \urlprefix\url{http://www.sciencedirect.com/science/article/pii/S0370157301000813}.

\bibitem[{\citenamefont{Blanter and B\"uttiker}(1998)}]{Blanter98}
\bibinfo{author}{\bibfnamefont{Y.~M.} \bibnamefont{Blanter}} \bibnamefont{and}
  \bibinfo{author}{\bibfnamefont{M.}~\bibnamefont{B\"uttiker}},
  \bibinfo{journal}{Phys. Rev. Lett.} \textbf{\bibinfo{volume}{81}},
  \bibinfo{pages}{4040} (\bibinfo{year}{1998}),
  \urlprefix\url{https://link.aps.org/doi/10.1103/PhysRevLett.81.4040}.

\bibitem[{\citenamefont{Fogedby and Imparato}(2012)}]{Fogedby12}
\bibinfo{author}{\bibfnamefont{H.~C.} \bibnamefont{Fogedby}} \bibnamefont{and}
  \bibinfo{author}{\bibfnamefont{A.}~\bibnamefont{Imparato}},
  \bibinfo{journal}{Journal of Statistical Mechanics: Theory and Experiment}
  \textbf{\bibinfo{volume}{2012}}, \bibinfo{pages}{P04005}
  (\bibinfo{year}{2012}),
  \urlprefix\url{http://stacks.iop.org/1742-5468/2012/i=04/a=P04005}.

\bibitem[{\citenamefont{Derrida et~al.}(1993)\citenamefont{Derrida, Evans,
  Hakim, and Pasquier}}]{Derrida1993}
\bibinfo{author}{\bibfnamefont{B.}~\bibnamefont{Derrida}},
  \bibinfo{author}{\bibfnamefont{M.~R.} \bibnamefont{Evans}},
  \bibinfo{author}{\bibfnamefont{V.}~\bibnamefont{Hakim}}, \bibnamefont{and}
  \bibinfo{author}{\bibfnamefont{V.}~\bibnamefont{Pasquier}},
  \bibinfo{journal}{Journal of Physics A: Mathematical and General}
  \textbf{\bibinfo{volume}{26}}, \bibinfo{pages}{1493} (\bibinfo{year}{1993}),
  \urlprefix\url{https://doi.org/10.1088%2F0305-4470%2F26%2F7%2F011}.

\bibitem[{\citenamefont{Sch{\"u}tz and Domany}(1993)}]{Schutz1993}
\bibinfo{author}{\bibfnamefont{G.}~\bibnamefont{Sch{\"u}tz}} \bibnamefont{and}
  \bibinfo{author}{\bibfnamefont{E.}~\bibnamefont{Domany}},
  \bibinfo{journal}{Journal of Statistical Physics}
  \textbf{\bibinfo{volume}{72}}, \bibinfo{pages}{277} (\bibinfo{year}{1993}),
  ISSN \bibinfo{issn}{1572-9613},
  \urlprefix\url{https://doi.org/10.1007/BF01048050}.

\bibitem[{\citenamefont{B\'erut
  et~al.}(2016{\natexlab{a}})\citenamefont{B\'erut, Imparato, Petrosyan, and
  Ciliberto}}]{Berut16}
\bibinfo{author}{\bibfnamefont{A.}~\bibnamefont{B\'erut}},
  \bibinfo{author}{\bibfnamefont{A.}~\bibnamefont{Imparato}},
  \bibinfo{author}{\bibfnamefont{A.}~\bibnamefont{Petrosyan}},
  \bibnamefont{and}
  \bibinfo{author}{\bibfnamefont{S.}~\bibnamefont{Ciliberto}},
  \bibinfo{journal}{Phys. Rev. Lett.} \textbf{\bibinfo{volume}{116}},
  \bibinfo{pages}{068301} (\bibinfo{year}{2016}{\natexlab{a}}),
  \urlprefix\url{https://link.aps.org/doi/10.1103/PhysRevLett.116.068301}.

\bibitem[{\citenamefont{B\'erut
  et~al.}(2016{\natexlab{b}})\citenamefont{B\'erut, Imparato, Petrosyan, and
  Ciliberto}}]{Berut16a}
\bibinfo{author}{\bibfnamefont{A.}~\bibnamefont{B\'erut}},
  \bibinfo{author}{\bibfnamefont{A.}~\bibnamefont{Imparato}},
  \bibinfo{author}{\bibfnamefont{A.}~\bibnamefont{Petrosyan}},
  \bibnamefont{and}
  \bibinfo{author}{\bibfnamefont{S.}~\bibnamefont{Ciliberto}},
  \bibinfo{journal}{Phys. Rev. E} \textbf{\bibinfo{volume}{94}},
  \bibinfo{pages}{052148} (\bibinfo{year}{2016}{\natexlab{b}}),
  \urlprefix\url{https://link.aps.org/doi/10.1103/PhysRevE.94.052148}.

\end{thebibliography}

\end{document}